\documentclass[twocolumn,showpacs,preprintnumbers,amsmath,amssymb,prb]{revtex4}

\usepackage{graphicx}
\usepackage{dcolumn}
\usepackage{longtable}
\usepackage{bm}
\usepackage{epsfig}

\begin{document}

\title{\textbf{Angular dependence of core hole screening in LiCoO$_2$: \\ A DFT+U calculation of the oxygen and cobalt \textit{K}-edge x-ray absorption spectra}}
\author{Am\'elie Juhin}
\email{A.F.Juhin@uu.nl} 
\affiliation{Inorganic Chemistry and Catalysis,~Department of Chemistry,~Utrecht University, Sorbonnelaan 16, 3584 CA Utrecht, Netherlands}
\author{Frank de Groot}
\affiliation{Inorganic Chemistry and Catalysis,~Department of Chemistry,~Utrecht University, Sorbonnelaan 16, 3584 CA Utrecht, Netherlands}
\author{Gy$\mathrm{\ddot{o}}$rgy Vank\'o}
\affiliation{KFKI Research Institute for Particle and Nuclear Physics, P. O. Box 49, H-1525 Budapest, Hungary}
\author{Matteo Calandra}
\affiliation{Institut de Min\'eralogie et de Physique des Milieux Condens\'es (IMPMC), UMR CNRS 7590,~Universit\'e Pierre et Marie Curie,~Universit\'e Paris Diderot, IRD UMR 206, IPGP, 4 place Jussieu, 75052 Paris Cedex 05, France}
\author{Christian Brouder}
\affiliation{Institut de Min\'eralogie et de Physique des Milieux Condens\'es (IMPMC), UMR CNRS 7590,~Universit\'e Pierre et Marie Curie,~Universit\'e Paris Diderot, IRD UMR 206, IPGP, 4 place Jussieu, 75052 Paris Cedex 05, France}

\begin{abstract}

Angular dependent core-hole screening effects have been found in the cobalt \textit{K}-edge x-ray absorption spectrum of LiCoO$_2$, using high-resolution data and parameter-free GGA+U calculations. The Co $1s$ core-hole on the absorber causes strong local attraction. The core-hole screening on the nearest neighbours cobalt induces a 2~eV shift in the density of states with respect to the on-site $1s-3d$ transitions, as detected in the Co \textit{K} pre-edge spectrum. Our DFT+U calculations reveal that the off-site screening is different in the out-of-plane direction, where a 3~eV shift is visible in both calculations and experiment. The detailed analysis of the inclusion of the core-hole potential and the Hubbard parameter U shows that the core-hole is essential for the off-site screening, while U improves the description of the angular dependent screening effects. In the case of oxygen \textit{K}-edge, both the core-hole potential and the Hubbard parameter improve the relative positions of the spectral features. 

\end{abstract}


\maketitle

\section{Introduction}

\begin{figure}[!b]
\includegraphics[width=7cm]{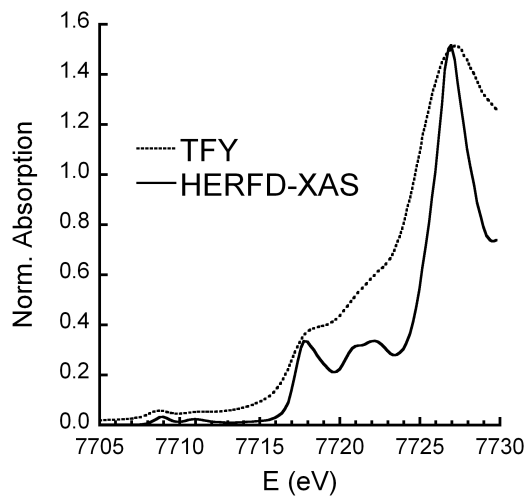}
\caption{\label{fig1} Comparison between XANES spectra recorded at the Co \textit{K}-edge using Total Fluorescence Yield (TFY, dashed line) and using High-Energy Resolution Fluorescence Detection (HERFD-XAS, solid line).}
\end{figure}

Layered LiCoO$_2$ has been extensively studied in relation to its wide use as a cathode material in rechargeable Li-batteries for portable devices. Upon electrochemical cycling (Li intercalation and deintercalation), defective cobaltites Li$_x$CoO$_2$ show fatigue and degradation, which limit the performance of the batteries. This has induced many experimental and theoretical efforts to investigate and interpret the modifications occuring in these materials.\cite{Marianetti,Hu,Uchimoto,Galakhov,Elp,Laubach,Petersburg,Ohzuku,Kramer} In particular, element - and orbital - selective spectroscopies, such as X-ray Absorption Spectroscopy, X-ray Emission Spectroscopy and X-ray Photoemision Spectroscopy, have been used to track local changes in the electronic and crystal structure, as a complement to X-Ray Diffraction. However, hardly any X-ray Absorption Near Edge Structure (XANES) data is available at the Co \textit{K}-edge in Li$_x$CoO$_2$ compounds, and all the existing data is measured in Total Fluorescence Yield, showing broad features, which are difficult to interpret.\cite{Kim,Alamgir,Poltavets} Over the past years, X-ray Absorption Spectroscopy using High Energy Resolution Fluorescence Detection (HERFD-XAS) has become a valuable tool to reveal subtle features of the absorption spectra, as can be direclty seen from Fig. \ref{fig1} at the Co \textit{K}-edge in LiCoO$_2$. HERFD-XAS is based on a two photons process (photon-in, photon-out) and thus may not yield true absorption spectra, due to strong electron-electron interactions in the final state. \cite{Carra} However, in the case of Co \textit{K}-edge in LiCoO$_2$, it has been shown that HERFD-XAS using the $K\alpha_1$ detection line yields the true absorption.\cite{Vanko} Thus, the considerably enhanced resolution gives new opportunities to infer unique details of the electronic structure of Co in LiCoO$_2$, which is the goal of this paper. 

Admittedly, the fine interpretation of the absorption features (i.e., beyond a 'finger print' analysis) requires parameter-free simulations of the spectra. First-principles calculations based on Density Functional Theory (DFT) have already proven to be very efficient for many systems, both to calculate the ground-state and the XANES spectra. However, a first limiting factor is the description of the electronic interaction, which is taken into account in a mean way (Local Density Approximation, labeled LDA, 
or General Gradient Approximation, labeled GGA). As an illustration, LiCoO$_2$ is known to be an insulator with a band gap of 2.7~$\pm$~0.3~eV,\cite{Elp} but LDA or GGA approaches underestimate the value of the gap in the ground state.\cite{Czizyk,Hu} To improve the description of the $3d$-electron correlation, a solution is to include the Hubbard parameter U.\cite{Anisimov} In the XANES calculation, an additional difficulty is to describe the core-hole effect. Both effects can possibly lead to a misassignement of the XANES features, due to wrong calculated peak positions and intensities. 
 
In this paper, we investigate the electronic structure and the XANES of LiCoO$_2$ using a parameter-free DFT+U approach, which has recently been developped. U is determined self-consistently as an intrinsic response of the material, and thus is no longer a fitting parameter.\cite{Cococcioni,Kulik} This method has been already applied to calculate the ground state of several systems (molecules and crystals), showing a significant improvement to describe the electronic structure.\cite{Kulik,Zhou,Hsu,Gougoussis,Gougoussis2} XANES calculations using DFT+U, though very promising, have only been performed on a restricted number of systems.\cite{Gougoussis,Gougoussis2} Here, we use these novel theoretical approaches to interpret finely the subtle XANES features revealed by HERFD-XAS at the Co \textit{K}-edge in LiCoO$_2$.

The paper is organized as follow. Section II gives the details of the XANES and of the Density Of States (DOS) calculations performed in LiCoO$_2$ using the GGA+U approach. Section III is devoted to the results and the discussion. First, we compare the ground state calculations obtained by GGA and GGA+U (Sec. IIIA). Then, after presenting the XANES spectra calculated at the Co \textit{K}-edge, we interpret the spectral features using local DOS and interpret the effect of U and of the core-hole on the theoretical spectrum (Sec. IIIB). In Sec. IIIC, we focus on the Co \textit{K} pre-edge region and interpret the origin of the local and non-local features. Finally, for completeness, we compare the theoretical XANES spectrum at the O \textit{K}-edge to experimental data recorded in Total Fluorescence Yield (Sec. IIID). Conclusions are given in Section IV. 

\section{Computational details}

First principles calculations were performed using the Quantum-Espresso first-principles total-energy code.\cite{PW}
The code uses plane waves and periodic boundary conditions. The XANES spectra are obtained in two steps: first the charge-density is obtained  self-consistently using the PW package of the Quantum-Espresso distribution,\cite{PW} then the XANES spectrum is computed in a continued fraction approach using the XSPECTRA package. \cite{Haydock,Haydock2,Gougoussis2}
As Co and O require large cutoffs to be simulated with standard norm-conserving pseudopotentials, we use ultrasoft pseudopotentials with two projectors per channel.\cite{Vanderbilt90} The electronic configurations are 3$d^8$ 4$s^1$ with non-linear core correction \cite{NLCC} for Co, 2$s^2$ 2$p^4$ for O and 2$s^1$ 2$p^0$ with non-linear core correction for Li. The use of Ultrasoft pseudopotentials limited the cutoff of the plane-wave expansion at 45~Ry energy. The generalized gradient approximation (GGA) was adopted. \cite{PBE}

When no core-hole was considered in the final state, a 12-atoms hexagonal unit cell with experimental lattice parameter and atomic positions was used.\cite{Akimoto} Electronic integration in the self-consistent run was performed on a 8$\times$8$\times$8 Monkhorst-Pack \textit{k}-point grid shifted along the \textit{c}-axis. The DOS and L$\mathrm{\ddot{o}}$wdin projected-DOS calculations were performed using a $12\times 12\times 12$ \textit{k}-points mesh and a Gaussian broadening of 0.3~eV.

Core-hole effects were treated in a supercell approach using a 3$\times$3$\times$1 hexagonal supercell containing 108 atoms (54 O, 27 Li, 26 Co and one absorbing Co) including the core-hole in the absorbing-atom pseudopotential. The distance between two neighboring absorbing atoms, upon the application of periodic boundary conditions, is 8.5~\AA, which is large enough to avoid interactions. In this case electronic integration in the self-consistent run was performed using a $4\times 4\times 4$ $k$-point Monkhorst-Pack mesh shifted along the \textit{c}-axis. The XANES calculation with a core-hole in the final states required electronic integration over a $4\times 4\times 4$ $k$-point grid. 

In the study of the electronic properties of LiCoO$_2$ the use of a Hubbard U parameter was found essential in order to reproduce the XAS spectrum and the magnitude of the electronic gap. In particular we impose U on the $d$-states of the Co atoms and calculate its value self-consistently (U$_{\mathrm{scf}}$) using the method of Refs \onlinecite{Cococcioni,Kulik}. We obtain U$_{\mathrm{scf}}$=5.6~eV. Thus there are no free parameters in our calculation. The XANES Co and O \textit{K}-edges cross-sections, both in their electric dipole (\textit{E1}) and electric quadrupole parts (\textit{E2}) were then calculated including the Hubbard U term. \cite{Gougoussis} The isotropic spectra were calculated in the correct point group symmetry ($\bar{3}m$ for Co, $3m$ for O) according to the procedure given in Appendix.
At the Co \textit{K}-edge, a constant broadening parameter (0.55~eV) was used in the continued fraction. This value is smaller than the broadening due to the Co $1s$ core-hole lifetime: indeed, the calculation is compared to experimental HERFD-XAS data recorded by monitoring the absorption using the \textit{K}$\alpha_1$ fluorescence line, which allows to remove partly the core-hole lifetime.\cite{Groot} The experimental data is taken from Ref. \onlinecite{Vanko}. For the O \textit{K}-edge, as the experimental data was measured in Total Fluorescence Yield, \cite{Elp} we used a broadening of 0.6~eV between 525~eV and 532~eV, and a broadening of 0.8~eV above 532~eV.

\section{Results and discussion}
\subsection{Ground state calculations}
 \begin{figure}[!t]
\includegraphics[width=8.5cm]{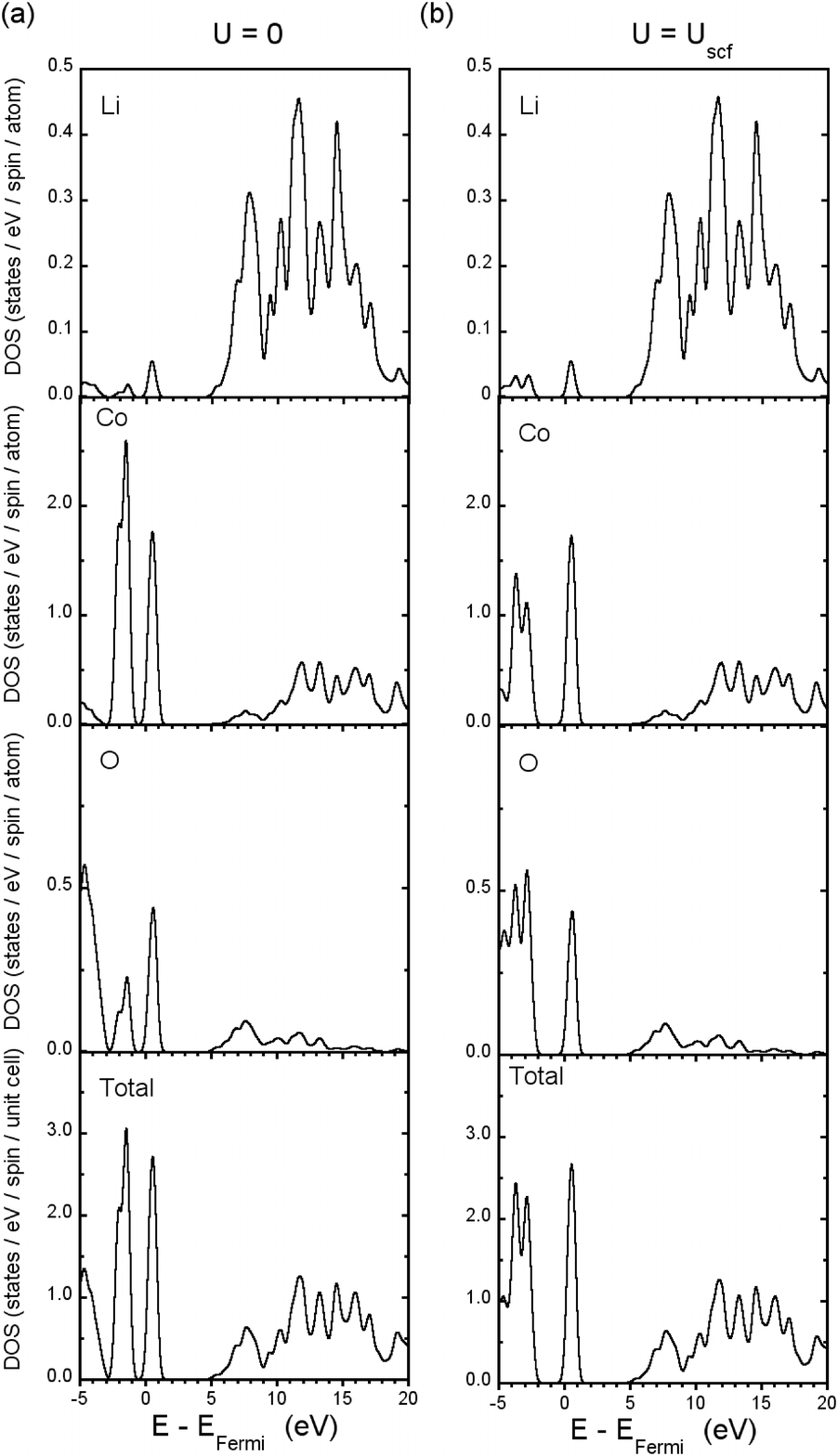}
\caption{\label{fig2} Density of states of LiCoO$_2$ calculated in GGA (pannel (a)) and GGA+U (pannel (b)), together with L$\mathrm{\ddot{o}}$wdin projected density of states on Li, Co and O atoms. The Fermi level is set at the bottom of the conduction band.}
\end{figure}

The total and L$\mathrm{\ddot{o}}$wdin projected DOS have been calculated in GGA and GGA+U for the ground-state of LiCoO$_2$ (i.e., with no core-hole). They are shown in Fig. \ref{fig2} (pannels (a) and (b), respectively). We used spin-polarization and started from an initial configuration, where each Co atom carries either a magnetic moment equal to zero, or slightly positive (0.2 $\mu_B$). In both cases, we observed a convergence to a non-magnetic state, where the final magnetic moment on the Co atoms is zero. This is consistent with previous experiments and calculations, which show that Co$^{3+}$ in LiCoO$_2$ is in a low spin state $S=0$. \cite{Bongers,Elp,Czizyk}
For U~=~0, DFT converges to an insulating ground state with a 1.1~eV electronic gap to be compared with the experimental
one 2.7~$\pm$~0.3~eV.\cite{Elp} The L$\mathrm{\ddot{o}}$wdin projected DOS show that the valence band is dominated by Co and O states, while the conduction band is dominated by Li and Co states. The gap is between $t_{2g}$ and $e_g$ Co states slightly hybridized with O $2p$ states. These results are consistent with previous DFT calculations on LiCoO$_2$. \cite{Hu,Czizyk}

DFT+U calculations using our calculated Hubbard parameter lead to a 2.3~eV gap (see panel (b) in Fig. \ref{fig2}),
in good agreement with the one determined from x-ray photoemission and bremsstrahlung isochromat spectroscopy.\cite{Elp} U shifts to lower energies the occupied states close to the Fermi level, i.e., Co $3d$ states plus Li $p$ and O $p$ states. For the empty states, the position in energy with respect to the Fermi level is identical to the calculation with U~=~0. 

\subsection{XANES at the Co \textit{K}-edge}

\begin{figure}[!b]
\includegraphics[width=8.5cm]{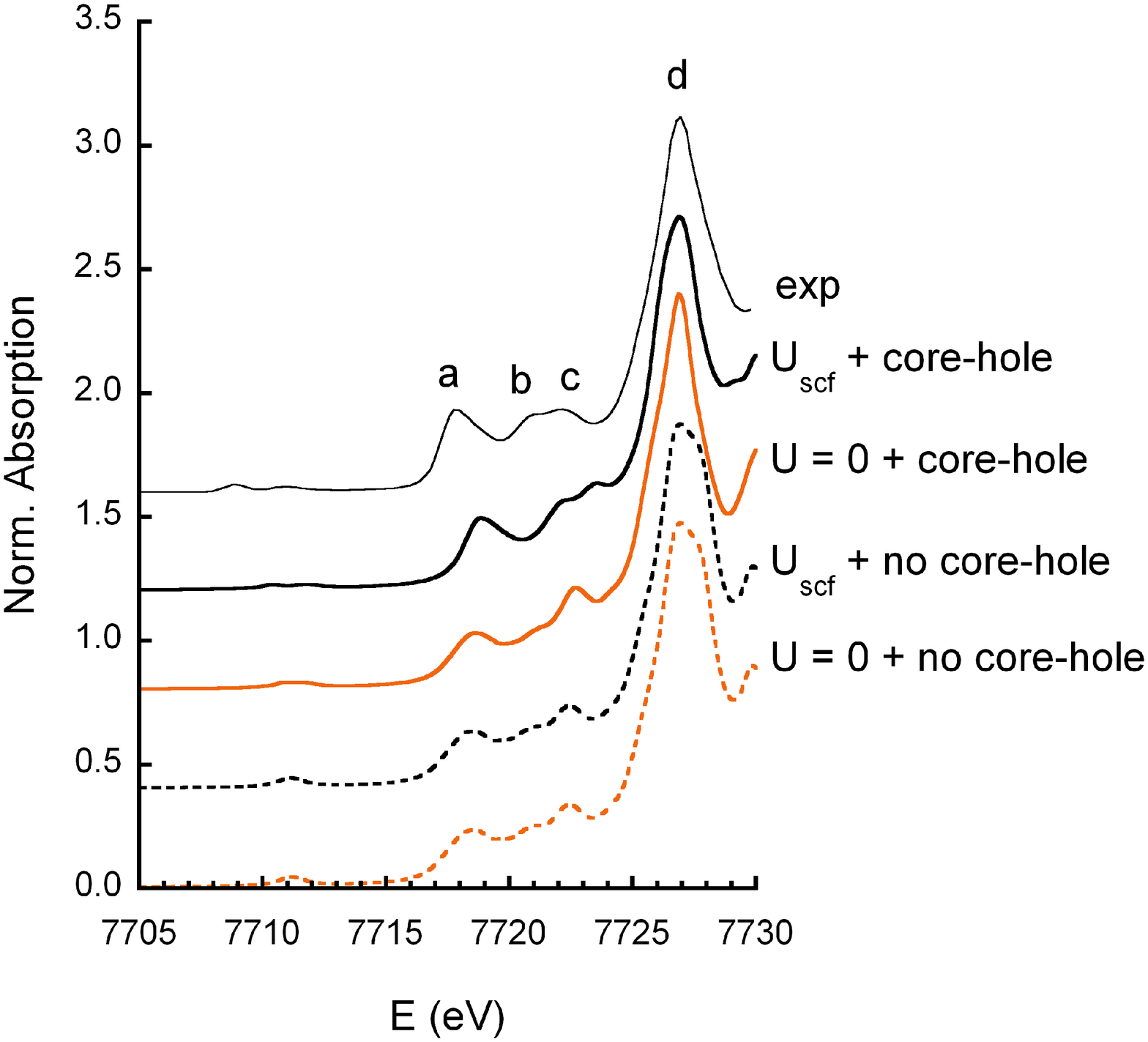}
\caption{\label{fig3} Comparison in the edge region between experimental (thin solid line) and theoretical isotropic XANES spectra at the Co \textit{K}-edge in LiCoO$_2$, obtained by four different calculations: GGA+U with or without the Co $1s$ core-hole (black and orange solid lines), GGA with or without the Co 1$s$ core-hole (black and orange dashed lines).}
\end{figure}

The calculated and experimental isotropic Co \textit{K}-edge spectra for LiCoO$_2$ are shown in Fig.~\ref{fig3}. 
The experimental HERFD-XAS spectrum (thin solid line) shows a set of three features at 7717.8~eV (peak a), 7720.9~eV (peak b) and 7722.2~eV (peak c), and the main edge is composed of a broad feature (peak d) centred at 7726.9~eV. This spectrum is compared to theoretical spectra obtained using four different GGA calculations, i.e., by switching on/off the Hubbard parameter U and on/off the Co $1s$ core-hole. These four spectra have been shifted in energy, so that the position of the main edge (maximum of the absorption) matches with the experimental one. 

First, all the experimental features are reproduced in the calculations. As mentioned in the introduction, computing the absorption cross-section is here fully justified, since at the Co \textit{K}-edge in LiCoO$_2$, the HERFD-XAS spectrum is a measurement of the true absorption.\cite{Vanko} By monitoring the influence of U and of the core-hole in the calculation, we can therefore draw conclusions on the core-hole effects in the material investigated experimentally. 

Second, when comparing the four theoretical spectra, differences in the peak positions and relative intensities are present. The best calculation is obtained when taking into account both the Hubbard parameter on the Co $3d$ orbitals and the Co $1s$ core-hole (black solid line). Indeed, the relative intensities and positions of peaks b and c, as well as the shape of peak d, are particularly improved. The Hubbard parameter has no effect when the core-hole is not included (orange and black dashed lines), since the relative positions of the peaks above the Fermi level are identical (see Fig. \ref{fig2}). The main effect of the core-hole for U~=~0 is to improve the shape of peak d, by merging the two subfeatures into a sharper one. When the core-hole is on, the Hubbard parameter mainly affects the intensity of peak a, the relative intensities and positions of peaks b and c, leading to a better agreement with experiment.  

\begin{figure}[!t]
\includegraphics[width=8.5cm]{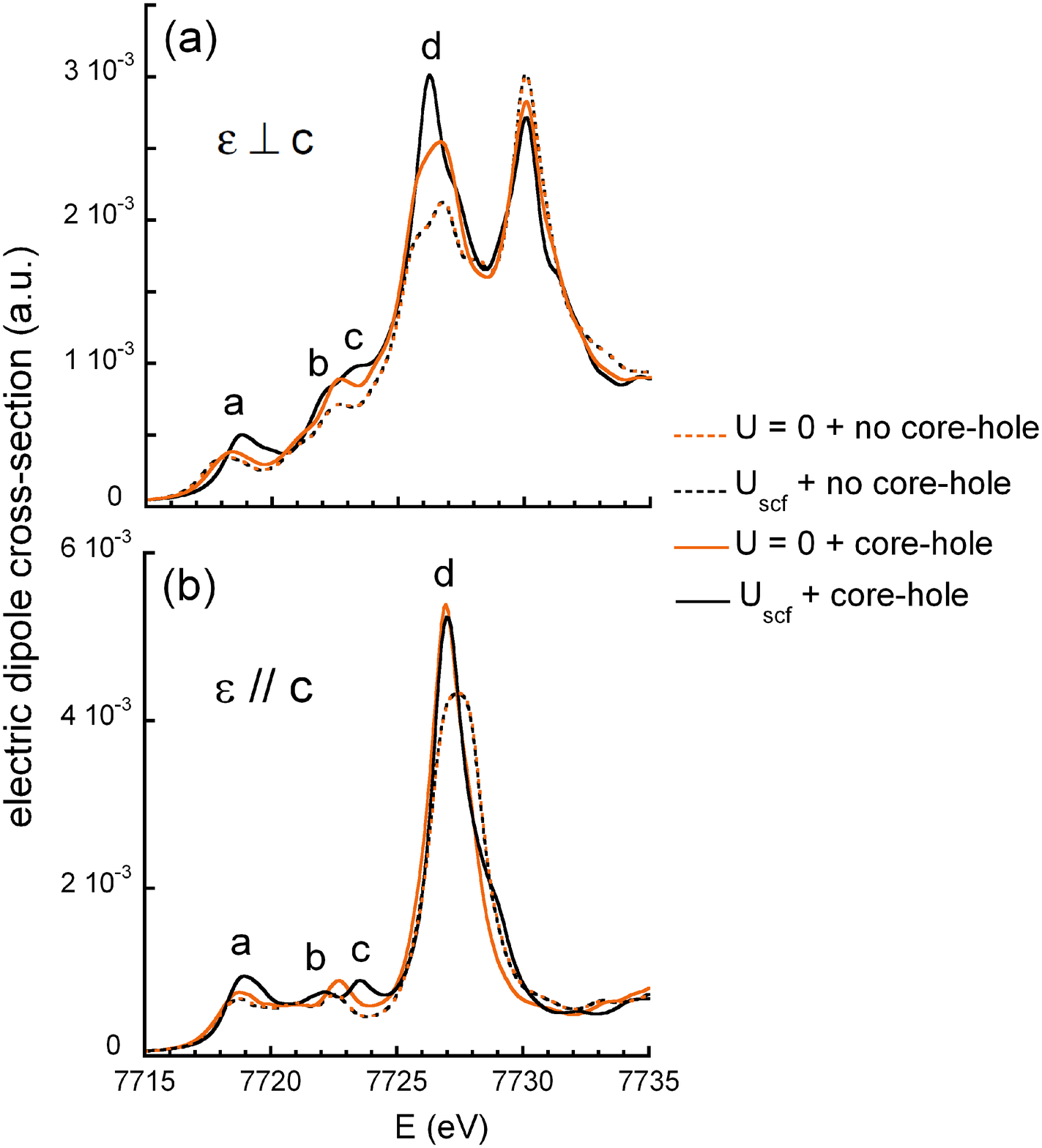}
\caption{\label{fig4} Electric dipole cross-section at the Co \textit{K}-edge calculated using the four different approaches. Pannels (a) and (b) show the cross-sections calculated for an incident polarization vector perpendicular and parallel to the \textit{c} axis.}
\end{figure}

In order to interpret the influence of U, of the core-hole and their possible anisotropic effects, we have calculated the electric dipole cross-sections for a polarization vector either perpendicular or parallel to the \textit{c} axis of the crystal (Figs. \ref{fig4}a and \ref{fig4}b). Indeed, as the spectra shown in Fig. \ref{fig3} are isotropic, they show an average absorption and thus they may be misleading. The dichroic XANES spectra ($\epsilon$ perpendicular and $\epsilon$ parallel) correspond respectively to transitions to the empty ($p_x$, $p_y$)- and $p_z$-states projected on the absorbing Co atom. Although no experimental data is available, the theoretical spectra enable a better understanding. In order to compare them, the spectra have been shifted in energy to match the position of the feature at 7730~eV (Fig. \ref{fig4}a). This feature corresponds indeed to delocalized Co $p$ states, which we assume to be less affected by U and the core-hole. U has no effect on the isotropic theoretical spectrum when the core-hole is not included. This can be explained by the fact that U is added on Co $3d$ orbitals, which are very localized states and do not contribute to the edge. When the Co $1s$ core-hole is added, the intensity of peak d is increased for both polarizations, together with a shift to lower energies. Both effects can be explained by the core-hole attraction and it affects in a similar way $p_x$, $p_y$ and $p_z$ states. This accounts for the fact that peak d becomes narrower in the isotropic spectrum, leading to a better agreement with experiment. When both the Hubbard parameter and the core-hole are included, peaks a, b and c are significantly affected. Peak a is shifted up and peak c is split into two subfeatures, which consequently makes peak b grow. As this was not present for U~=~0 with the core-hole, it can be attributed to the Hubbard parameter. These effects are similar for both polarizations. However, this is different for peak d, which is shifted to low energies and which is more intense, but only for the corresponding $p_z$ states. From the previous analysis, we can conclude that the Co $1s$ core-hole mainly affects the intensity of the XANES features by attracting Co empty $p$ states to lower energy. The Hubbard parameter on Co $3d$ states affects the peak positions, but only when the core-hole is present. This can be explained by the fact that the core-hole attraction mixes Co and O empty states in such a way that U can act also on the Co $p$ empty states.  

\begin{figure}[!t]
\includegraphics[width=8.5cm]{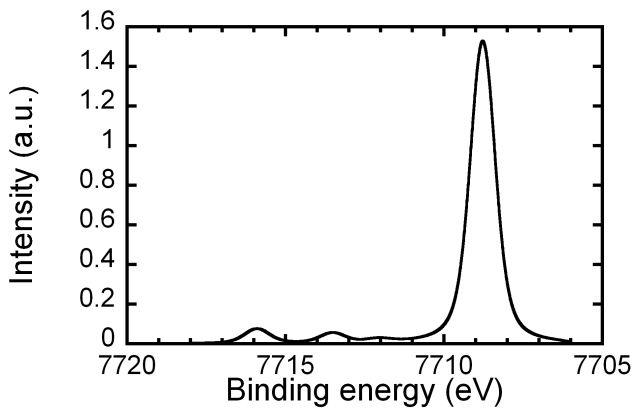}
\caption{\label{fig5} Theoretical Co $1s$ XPS cross-section calculated using the charge-transfer multiplet approach.}
\end{figure}

Finally, we want to comment on the fact that our DFT calculations reproduce all the experimental XANES features. Indeed, DFT can only take into account one single Slater determinant and thus neglects charge-transfer effects. When the latter are significant, additional features are visible in the edge of the experimental spectrum (e.g., Cu $K$-edge in La$_2$CuO$_4$)\cite{Tolentino}, but they cannot be calculated using a single-particle approach. The existence of charge transfer induced transitions can be inferred from $1s$ X-ray Photoemission Spectroscopy (XPS) spectra. If intense satellite features are visible, one expects a modification of the XANES $K$-edge features. The good agreement obtained between theoretical and experimental XANES spectra in LiCoO$_2$ enables already to conclude that charge transfer satellite transitions are negligible. To support this result, we have calculated the Co $1s$ XPS cross-section using the charge-transfer multiplet theory. This calculation used the values of the crystal field, the Slater integrals and the charge transfer parameters from Ref. \onlinecite{Elp}. It included two configurations in the initial and final states of XPS: $3d^6$ and $3d^7$\underline{L}, where \underline{L} denotes a hole on an oxygen ligand. The theoretical Co $1s$ XPS cross-section is shown in Fig. \ref{fig5}. We found that 95~\% of the intensity goes to the main peak, which is visible at 7709~eV in binding energy and which is the so-called well-screened peak, a mixture of mainly $1s^13d^7$\underline{L} with some $1s^13d^6$ final states. As the high-binding energy satellite features are very weak, satellite transitions in the XANES will not be visible, which is in line with our results.

\subsection{Insight from the Co \textit{K} pre-edge}

\begin{figure}[!t]
\includegraphics[width=8.5cm]{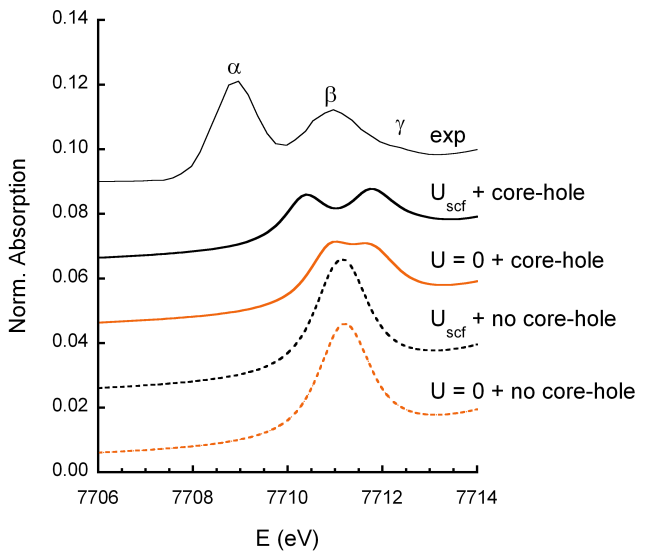}
\caption{\label{fig6} Comparison in the Co \textit{K} pre-edge region between experimental (thin solid line) and theoretical isotropic XANES spectra in LiCoO$_2$, obtained by four different calculations: GGA+U with or without the Co $1s$ core-hole (black and orange solid lines), GGA with or without the Co $1s$ core-hole (black and orange dashed lines).  }
\end{figure}

\subsubsection{Comparison between experiment and calculations}
We now do a similar comparative analysis in the Co \textit{K} pre-edge region (shown in Fig.~\ref{fig6}). The experimental isotropic pre-edge spectra (thin solid line) shows two main features, labeled $\alpha$ and $\beta$, which are visible respectively at 7708.9~eV and 7710.9~eV. Additionnally, a small shoulder labeled $\gamma$ is visible at 7712.3~eV. At this stage of the paper, we will focus the analysis on the intense peaks $\alpha$ and $\beta$, since peak $\gamma$ is not discriminative of the different calculations (the broadening used in the calculation being too large). The origin of this shoulder will be discussed later in the paper. 

All the theoretical spectra are shifted to high energy compared to experiment (the position of the edge being the same). Such a shift is systematically observed at the \textit{K} pre-edge of transition metal ions:\cite{Gougoussis,Juhin,Bordage,Cabaret} it is due to the fact that the $1s$ core-hole is overscreened in the calculation, which consequently underestimates the splitting between the electric dipole and electric quadrupole transitions. Nevertheless, as this shift is observed for all the theoretical spectra, we will now compare the relative intensities and peak positions. There are significant differences between the four theoretical spectra. When the Hubbard parameter and the core-hole are not included in the calculation (orange dashed line), the pre-edge consists of one single broad peak at 7711.1~eV. When U is added (orange dashed line), this feature is shifted only by 0.1~eV to lower energies. If the core-hole is taken into account, either with U~=~0 or with U~=~U$_{\mathrm{scf}}$ (orange and black solid lines), this single broad intense feature is split into two sharp ones with lower intensity. The splitting is higher for U~=~U$_{\mathrm{scf}}$, which yields the best agreement obtained between calculation and experiment. We can already conclude that for the Co \textit{K} pre-edge region, the core-hole affects the splitting of the two main peaks, and U influences strongly the splitting of the two, as well as their positions with respect to the edge. For U~=~U$_{\mathrm{scf}}$ and the core-hole, the intensity of peak $\alpha$ is however underestimated by 40~\% with respect to experiment, while that of peak $\beta$ is well reproduced. The splitting between peaks $\alpha$ and $\beta$ is also smaller than in the experimental data (1.4~eV vs 2.0~eV). We will further comment these points after analyzing in detail the origin of the pre-edge features. 

\begin{figure}[!t]
\includegraphics[width=8cm]{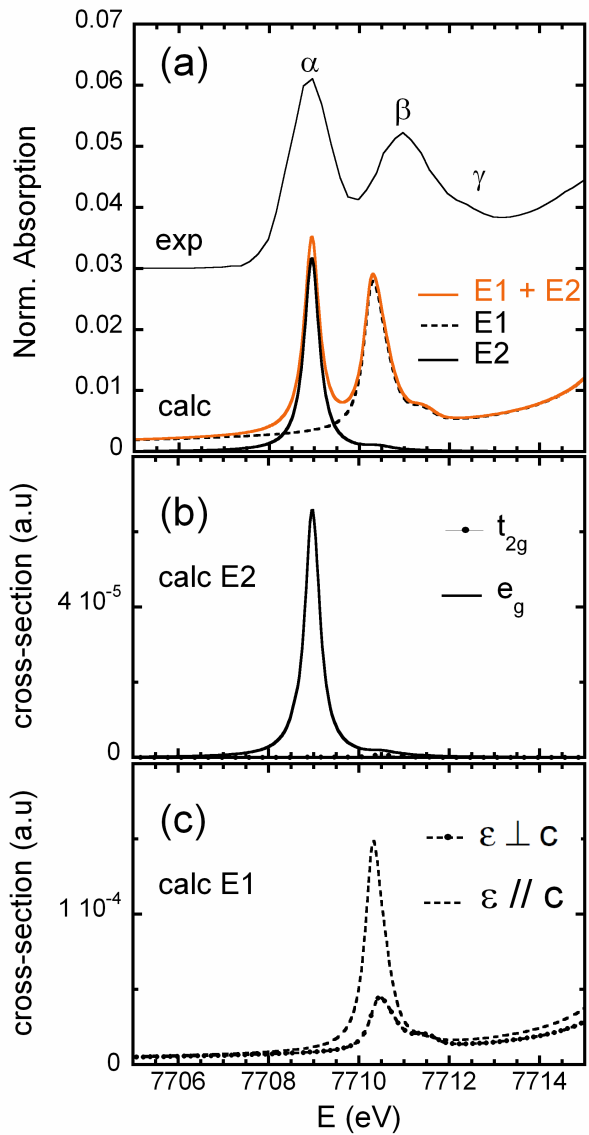}
\caption{\label{fig7} (a) Comparison between experimental (thin solid line) and theoretical isotropic pre-edge spectra (solid line) at the Co \textit{K}-edge in LiCoO$_2$. For the theoretical spectrum we show the electric dipole contribution (\textit{E1}, dashed line), the electric quadrupole contribution (\textit{E2}, solid line) and their sum (\textit{E1}+\textit{E2}, orange solid line). (b) Angular dependence of \textit{E2} cross-section. (c) Angular dependence of \textit{E1} cross-section. The theoretical spectrum has been shifted to the experimental data in order to match the position of peak $\alpha$, for a more direct comparison of the peak position.}
\end{figure}

The origin of the pre-edge features can be clarified by plotting separately the electric dipole (\textit{E1}) and the electric quadrupole (\textit{E2}) contributions. Figure \ref{fig7}a shows the \textit{E1} (solid line) and \textit{E2} (dashed line) contributions, as well as the sum of the two (orange solid line), for the calculation performed in GGA+U and with the core-hole. The theoretical spectrum has been shifted to the experimental data in order to match the position of peak $\alpha$, for a more direct comparison of the peak position. Note that in this case, the calculation has been performed with a smaller broadening parameter (0.2~eV) than in Fig. \ref{fig6}, in order to separate clearly the three peaks. In particular, it improves visibly the intensity ratio between the peaks. Therefore, only the relative intensities should be compared to the experimental ones. From Fig. \ref{fig7}a, we can conclude that peak $\alpha$ is due to pure electric quadrupole transitions ($1s-3d$) and that peak $\beta$ is due to almost pure electric dipole transitions ($1s-p$). Peak $\gamma$ is due to pure electric dipole transitions. 

\begin{figure*}[t!]
\includegraphics[width=18cm]{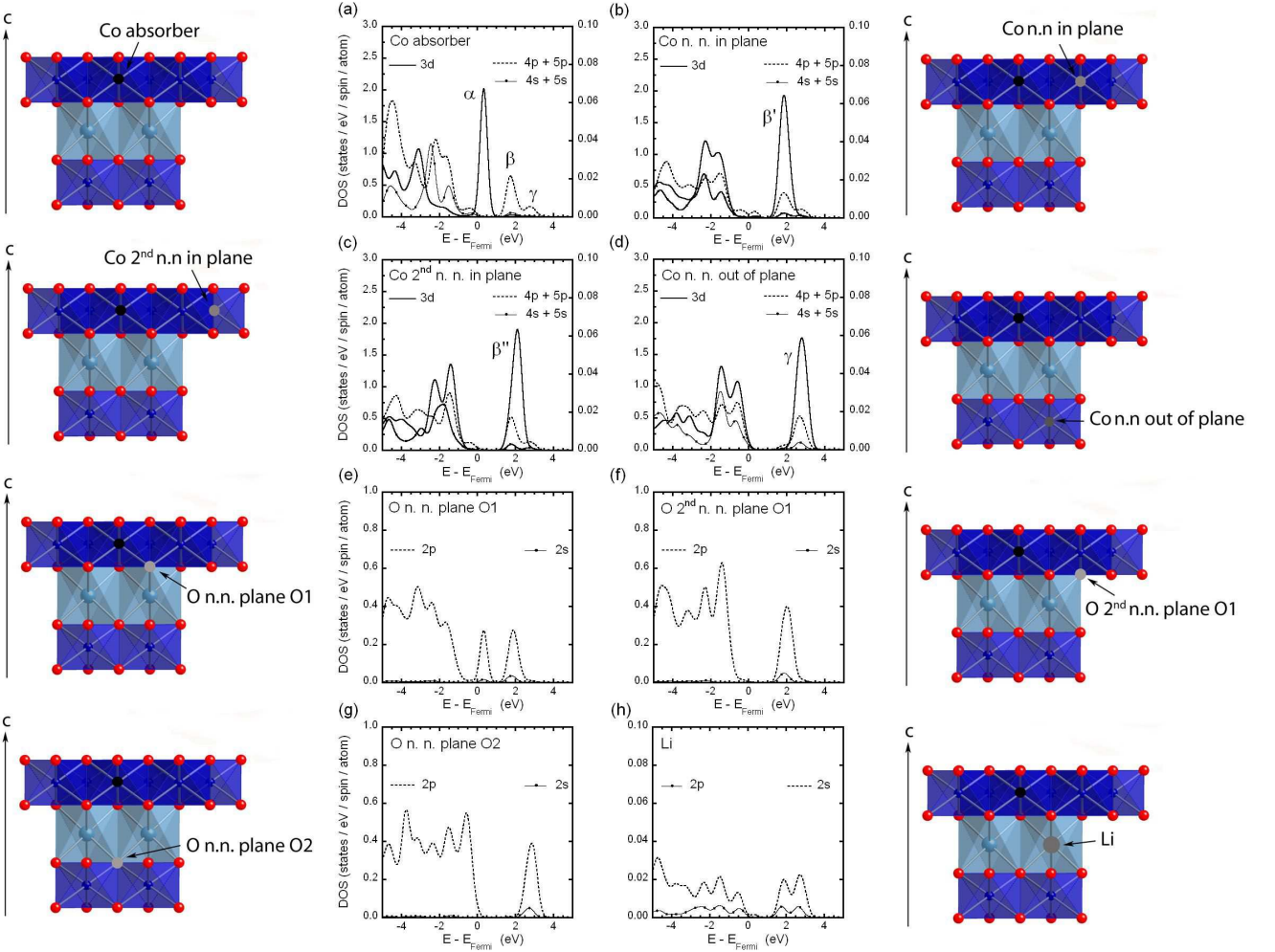}
\caption{\label{fig8} Density of states of a LiCoO$_2$ supercell including a Co $1s$ core-hole and U~=~U$_{\mathrm{scf}}$. L$\mathrm{\ddot{o}}$wdin projected density of states are plotted for the absorbing Co atom and its the nearest Li, O and Co neighbors. The $s$, $p$ and $d$ DOSs are given (solid line with circles, dashed and solid lines, respectively).}
\end{figure*}

To go further into the analysis, we have calculated the angular dependence of the electric dipole and the electric quadrupole cross-sections. This can be achieved by calculating the absorption cross-section for various orientations of the polarization and wave vectors, with respect to the crystal. Figure \ref{fig7}b shows that peak $\alpha$ corresponds to $3d$ states that have, non-surprisingly, a pure $e_g$ character. Indeed, when the transition operator $\bm{\mathrm{\hat{\varepsilon}}}\cdot\textbf{r}\bm{\mathrm{\hat{k}}}\cdot\textbf{r}$ is chosen along a Co-O bond (i.e., $e_g$-like orbitals are probed), the intensity of peak $\alpha$ is maximized. When it is chosen between two Co-O bonds (i.e., $t_{2g}$-like orbitals are probed), peak $\alpha$ disappears almost completely (Fig. \ref{fig7}b). Such X-ray Linear Natural Dichroism effects are already well-known and are used to derive local information on the electronic and crystallographic structure.\cite{Juhin} However, in the case of Co$^{3+}$, they are particularly emphasized, since the $e_g$ orbitals are the only ones to be empty. Because the Co site point group symmetry is not $O_h$ but slightly lower ($D_{3d}$), it is not possible, in principle, to extinct completely peak $\alpha$ because of a slight hybridization between $e_g$ and $t_{2g}$ states. However, we found out that peak $\alpha$ has almost a pure $e_g$ character, which shows that the distortion of the Co site from perfect octahedral symmetry is very small. 
Figure \ref{fig7}c shows that peak $\beta$ has a mixed $p_x$, $p_y$, $p_z$ character and the $p_z$ states are at slightly lower energy. Peak $\gamma$ has a dominant ($p_x$,$p_y$) character: its intensity can thus be enhanced by doing measurement on a single crystal using a polarization vector perpendicular to the \textit{c} axis. 

\subsubsection{Assignement of the pre-edge features}
To assign these transitions, the DOS for the supercell including the Co \textit{1s} core-hole and U~=~U$_{\mathrm{scf}}$ has been projected on the absorbing Co atom and selected neighboring atoms (O, Li and Co) in a range of 5~\AA~from the absorber. The local DOSs are plotted in Fig.\ref{fig8}. We distinguish: the first and second oxygen neighbors (O in \textit{O1 plane} at a distance of 1.92~\AA~and 3.41~\AA~respectively), the first and second Co neighbors in the same plane as the absorber (Co \textit{in plane}, at a distance of 2.82~\AA~and 4.88~\AA~respectively), the first Li neighbors (\textit{Li plane}, at a distance of 2.85~\AA), the first Co neighbors in the plane below the absorber (Co \textit{out of plane}, at a distance of 4.96~\AA), and the first O neighbors between the \textit{Li plane} and the Co \textit{out of plane} (O in \textit{O2 plane}, at a distance of 3.66~\AA).

As can be seen from the local $d$-DOS on the absorbing Co (Fig. \ref{fig8}a, solid line), peak $\alpha$ is due to electric quadrupole transitions from the $1s$ orbital to $3d$ states localized on the Co absorber (intrasite excitations). Moreover, although empty $2p$ states of the O nearest neighbors are found at the same energy position (Fig. \ref{fig8}e, dashed line), we found no evidence of direct transitions to these states, contrary to the case of Ni \textit{K} pre-edge in NiO. \cite{Gougoussis} 

The local $p$-DOS on the absorbing Co atom (Fig. \ref{fig8}a, dashed line) shows that peak $\beta$ is mainly due to electric dipole transitions, i.e. from the $1s$ orbital to $4p$ empty states. At this energy (2~eV above the Fermi level), there are actually two different contributions slightly shifted one from each other: one centred at 1.6~eV (peak $\beta'$) and a second one, centred at 2.0~eV (peak $\beta''$). 
This accounts for the fact that peak $\beta$ is broader than peak $\alpha$. The local DOS on the neighbors lying in the same plane as the absorber clarify the origin of these two subfeatures. Peak $\beta'$ is due to transitions to empty $4p$ states of the absorber, hybridized with $2p$ states of the O nearest neighbors (Fig. \ref{fig8}e, dashed line) and $3d$ states of the Co nearest neighbors (Fig. \ref{fig8}b, solid line). Peak $\beta''$ is due to transitions to empty $4p$ states of the absorber, hybridized mainly with $2p$ states of the O second-nearest neighbors (Fig. \ref{fig8}f, dashed line) and with $3d$ states of the Co second nearest neighbors (Fig. \ref{fig8}c, solid line). The splitting between peaks $\beta'$ and $\beta''$ is due to a different core-hole screening between the (Co,O) first neighbors and the (Co,O) second neighbors: as the latter lie further from the absorber, the Co $1s$ core-hole is more screened. Therefore, peak $\beta$ has a double non-local (\textit{off-site}) origin, which is revealed by the core-hole attraction. We have found that peaks $\beta'$ and $\beta''$ both owe their broadening to the fact that the Co $4p_z$ empty states lie at lower energies than the ($4p_x$, $4p_y$) states, as shown in Fig. \ref{fig7}c. However, this is not due to a different screening of the core-hole along these directions, since in the ground-state (i.e., with no core-hole), the Co $p_z$ empty states already lie lower in energy (0.2~eV) than the empty $p_x,p_y$ states. Therefore, it is the strong anisotropy of the structure that introduces direction-dependent hybridizations and lifts orbital degeneracy, even when the core-hole is absent. 

Peak $\gamma$ is due to non-local excitations, i.e., to \textit{on-site} Co empty $4p$ states hybridized with empty $3d$ states of the Co atoms, which are in the plane below and above the absorber (Co \textit{out of plane}, Fig. \ref{fig8}d). Such \textit{far off-site} transitions (at 2.6~eV above the Fermi level) are mediated by the Li $2p$ empty orbitals (Fig. \ref{fig7}h, dashed line) and the $2p$ states of the O atoms (nearest neighbors \textit{plane} O2, dashed line), which are between the Li and the Co \textit{out of plane} atoms. These Co atoms lie a distance of 4.96~\AA~from the absorber, which is close to the distance between the absorber and the Co second neighbors \textit{in plane}. This implies that the distance being the same, the core-hole is more screened along the \textit{c} direction than in the perpendicular plane. As a consequence, the splitting between peaks $\beta$ and $\gamma$ is due to an angular dependent screening of the core-hole. This original result can certainly be attributed due to the anisotropic layered structure of LiCoO$_2$. As we mentioned before, the far off-site peak has mainly a ($p_x$, $p_y$) character. It can appear at first sight surprising that out of plane hybridization is achieved mainly by the O in plane $p$-orbitals, taking into account the strong anisotropy of the structure. However, a look at the structure reveals that the $c$ (or $z$) axis is directed along the local $C_3$ axis of the CoO$_6$ octahedra. This implies that the Co-O bonds are not directed along the $x$, $y$, and $z$ directions and that the $p_x$, $p_y$ or $p_z$ character discussed does not correspond to the traditionnal reference frame. In such a description, it is thus not surprising to have a $p_x$, $p_y$ character in the peak probing the Co atoms out of plane.  

\subsubsection{Influence of U and of the core-hole}

As mentioned before, both the Hubbard parameter and the core-hole must be included in the calculation of the Co \textit{K} pre-edge spectrum, in order to obtain a good agreement with experiment. The different features are well separated and this is due to a complex combined effect. 
When the core-hole is off, we could expect that U shifts the empty Co $3d$ states up towards the edge, since the latter consists of Co $p$ states not affected by U. This is however not the case. When the core-hole is switched on for U~=~0, peaks $\alpha$, $\beta$ and $\gamma$ are separated by the core-hole attraction. This different behavior is due to the non-local character of peaks $\beta$ and $\gamma$.
The splitting between $\alpha$ and $\beta$ increases, since the screening is all the more efficient that the Co and O neighbors are far from the absorber. The energy difference between $\beta$ and $\gamma$ increases as well, because core-hole effects show an angular dependence (in plane and out of plane). 

When the core-hole is present, the Hubbard parameter on Co $3d$ orbitals increases again the energy difference between the in-site and the off-site peaks, on one hand, and between the two off-site peaks, on the other hand. This is similar to the case of Ni \textit{K}-edge in NiO.\cite{Gougoussis} U has an effect since the $3d$ empty states of the absorber and of its neighbors have already been separated by the differential core-hole screening. This is in line with the fact that U has hardly no effect if the core-hole is absent. 

The fact that the intensity of peak $\alpha$ is slightly underestimated compared to the two other ones raises the question of a possible missing electric dipole contribution at this energy. This can have two possible origins: (i) the quality of the sample investigated experimentally, (ii) vibrational effects - missing in the calculation-, which introduce an electric dipole contribution in peak $\alpha$. First, the sample is finely grinded. As most powders remain textured and the crystal structure is trigonal, there could be angular dependence effects. However, these effects should be rather small, and we do not think this could explain the observed disagreement. We recall that the calculated spectrum is the isotropic one, and thus should be directly compared to the experiment. Second, another possibility is the vibrational effects, since the experimental data was measured at room temperature. At T~=~0~K, the fact that the Co site is centrosymmetric ($D_{3d}$ point group) prevents any static mixing between the Co \textit{3d} states and the O $2p$ states in the calculated XANES spectrum. However, lattice vibrations can remove this rule and thus, significant changes in the XANES spectrum may be possible. 
In the case of Co \textit{K}-edge in LiCoO$_2$, vibrations could introduce an electric dipole contribution in peak $\alpha$. To verify the validity of this conjecture, HERFD-XAS measurements at the Co \textit{K}-edge as a function of temperature should be performed and combined with a consequent analysis including the vibrational effects.\cite{Brouder} 

\subsection{XANES at the O \textit{K}-edge}

\begin{figure}[!b]
\includegraphics[width=8cm]{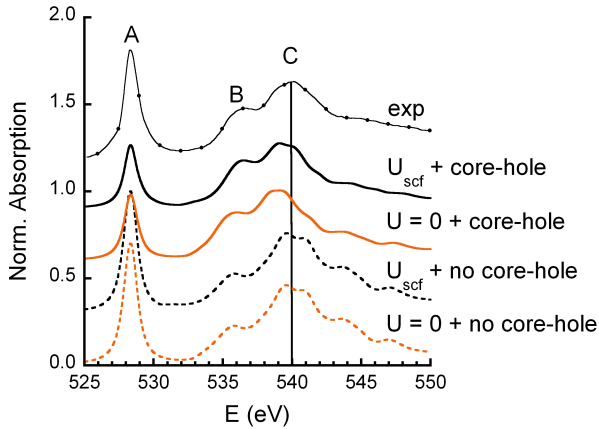}
\caption{\label{fig9} Comparison between experimental (solid line with circles, taken from Ref. \onlinecite{Groot}) and theoretical isotropic XANES spectra at the O \textit{K}-edge in LiCoO$_2$, obtained by two different approximations, GGA+U and GGA (black and orange), both including the O $1s$ core-hole (solid and dashed lines).}
\end{figure}

For completeness, we have calculated in DFT+U the XANES at the O \textit{K}-edge, although the experimental data has been measured in Total Fluorescence Yield.\cite{Elp} 
The experimental isotropic O \textit{K}-edge spectrum is shown in Fig. \ref{fig9} (solid line with filled circles). It shows a well-defined feature at 528.3~eV (peak A) and two broad ones at 536.4~eV (peak B) and 539.9~eV (peak C). It is compared with the theoretical ones calculated in the electric dipole approximation (i) in GGA+U with or without the O $1s$ core-hole (black and orange solid lines, respectively) and (ii) in GGA with or without the O $1s$ core-hole (black and orange dashed lines, respectively). The four theoretical spectra have been shifted in energy in order to match the position of peak A with the experimental data. 

\begin{figure}[!]
\includegraphics[width=9cm]{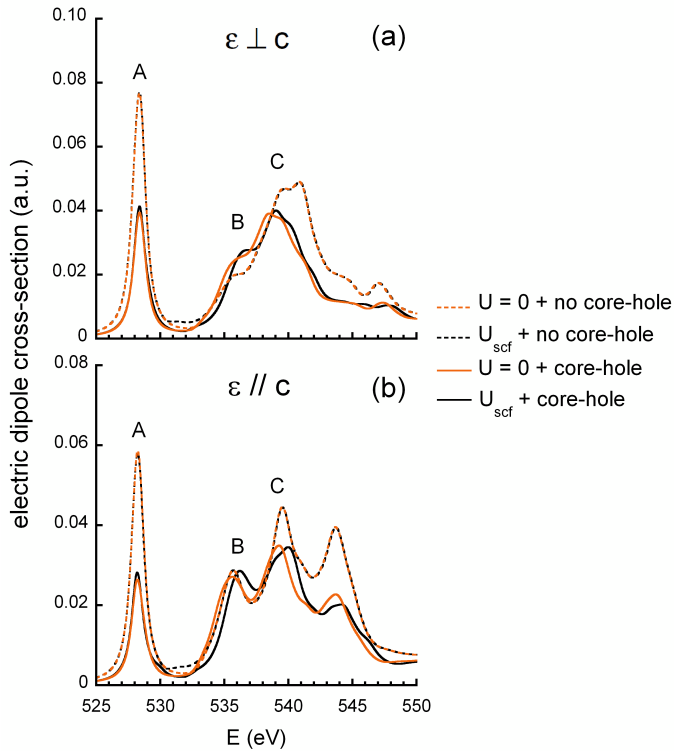}
\caption{\label{fig10} Electric dipole cross-section at the O \textit{K}-edge calculated using the four different approaches. Pannels (a) and (b) show the cross-sections calculated for an incident polarization vector perpendicular and parallel to the \textit{c} axis.}
\end{figure}

The best agreement has been obtained for the calculation using U~=~U$_{\mathrm{scf}}$ and the core-hole (black solid line): although the intensity of peak A is underestimated, the relative position of the peaks is improved compare to the calculations without the core-hole or without the Hubbard parameter. This is also the case for the shape of peak C and the intensity ratio between peaks B and C. 

The main effect of the core-hole is to decrease by a factor of two the intensity of peak A, which corresponds to transitions to empty O $2p$ states. This is due to the fact that part of these states are shifted down to -10~eV below the Fermi level (not shown), due to a strong underestimation of the core-hole screening. Similar overestimation of core-hole effects has also been reported in the calculations performed at the \textit{K}-edge of light elements such as B and O,\cite{Jiang,Hetenyi} where it was found that a half core-hole provides a better agreement in the region just above the edge. Such a drastic influence has also been shown at the Cu \textit{K}-edge {\bf \cite{Gougoussis2}}, but only for the electric quadupole cross-section. Here, we show that the core-hole effects are not well described also for the electric dipole cross-section, since the intensity of peak A is better reproduced without the core-hole (though slightly overestimated). 
The core-hole improves the shape of peak C and the intensity ratio between peaks B and C, which can be understood by plotting the electric dipole cross-sections for a polarization vector $\epsilon$ either perpendicular or parallel to the \textit{c} axis (Figs. \ref{fig10}a and \ref{fig10}b). The attraction of the core-hole induces a shift by -1.2~eV of the ($p_x$, $p_y$) states at 540~eV, and an overall decrease in intensity of the corresponding $p$ states. Part of the $p_z$ states from peak C have been transfered to peak B (536~eV). For $p_x$, $p_y$ states, it is likely that the core-hole shifts them below the Fermi level, since peak B has the same intensity with and without the core-hole and the intensity of peak C decreases. 

The Hubbard parameter U has no effect on the theoretical spectrum when the core-hole is absent, which is in line with the empty DOS shown in Fig.\ref{fig2}. However, it does have an effect when the core-hole is on and affects the peak positions, similarly to the case of Co \textit{K}-edge. Although U is added on the Co $3d$ orbitals, it has an effect on the empty O $p$ states because they are slightly hybridized (Fig.\ref{fig2}) and because the core-hole introduces differential attraction on the O empty states. This results in a shift to higher energies of peaks B and C relatively to peak A.

\section{Conclusions}

For the first time, the XANES spectra at the Co \textit{K}-edge and O \textit{K}-edge have been calculated in LiCoO$_2$ using a parameter-free GGA+U approach. A good agreement has been obtained with the experimental data, which has enabled to draw conlusions of several types.

First, from the computational point of view, we have shown that the Hubbard parameter U improves the theoretical spectra in the edge region, for both the relative intensities and positions of the spectral features. This improvement is even more striking at the Co \textit{K} pre-edge, where both the core-hole and U are needed to interpret the spectral features. The core-hole effects were found to be the factor preventing a better agreement with experiment: in the calculation, where we considered a static core-hole, the O $1s$ core-hole was found to be too attractive, while the screening of the Co $1s$ core-hole is overestimated. A better treatment of the core-hole effects could be achieved using the Bethe-Salpeter equation, which treats electron and hole dynamics \textit{ab initio}, as well as electron-hole interactions.\cite{Shirley}

Second, we have shown that the GGA+U approach improves significantly the description of the electronic structure for the ground state in LiCoO$_2$, since the value of the theoretical electronic band gap is, for the first time, consistent with the experimental one. This implies that this approach could yield better results on the electronic structure of Li-defective cobaltites Li$_x$CoO$_2$. 

Third, high-resolution XANES data at the Co \textit{K} pre-edge has revealed new subtle features, which have been interpreted thanks to DFT+U calculations. In the pre-edge, in addition to classical on-site $1s$-$3d$ transitions, we have shown the existence of two different Co~$4p$~-~Co~$3d$ intersite hybridizations, in plane and out of plane. This is achieved via a strong Co~$3d$~-~O~$2p$ hybridization in the Co planes, and via a strong Co~$3d$~-~O~$2p$~-~Li $2p$ hybridization in the direction perpendicular to the Co planes. Although it can be argued that the presence of the core-hole prevents to draw any conclusion on the ground-state properties, we want to point out that these pre-edge features can only be resolved because the Co $1s$ core-hole is present: the screening of the core-hole is indeed dependent both on the distance and on the direction. This means that HERFD-XAS measurements performed on a single crystal, coupled to GGA+U parameter-free calculations, can be used to track fine changes in orbital hybridization, such as the ones induced by electrochemical cycling in the electronic and crystallographic structure of Li$_x$CoO$_2$ compounds. For example, as the increase in Li-vacancies leads to the augmentation of the $c$ parameter,\cite{Laubach,Reimers} it should affect the relative intensities and peak positions of the Co pre-edge features.

\begin{acknowledgments}
The authors are grateful to D. Cabaret, M. Lazzeri and C. Gougoussis for fruitful discussions. The theoretical part of this work was supported by the French CNRS computational Institut of Orsay (Institut du D\'eveloppement et de Recherche en Informatique Scientifique) under project 92015. G.V. acknowledges financial support from the Hungarian Scientific Research Fund (OTKA) under contract No. K72597 and from the Bolyai Fellowship.
\end{acknowledgments}

\newpage

\section{Appendix}

\subsection{Calculation of the isotropic electric dipole cross-section}
For a general symmetry, the isotropic electric dipole cross-section is 
given by $(\sigma_{xx}+\sigma_{yy}+\sigma_{zz})/3$, where
$\sigma_{xx}$, $\sigma_{yy}$ and $\sigma_{zz}$ are the
electric dipole absorption spectra calculated along three
perpendicular directions~\cite{ChB90}.
We now consider the sites of oxygen and cobalt.

In the space group $R\overline{3}m$ described in the hexagonal setting, 
there are two equivalent oxygen atoms in position \textit{6c}: 
(0 0 $z$) and (0 0 -$z$). The four other O sites can be obtained by 
applying the lattice translations (1/3, 2/3, 2/3) and (2/3, 1/3, 1/3). 
The point group symmetry is \textit{3m} (or $C_{3v}$). 
Therefore, we take $z$ along the $\mathbf{c}$ axis of the crystal
and the local symmetry group gives us 
$\sigma_{xx}=\sigma_{yy}$.

The Co atom is in position \textit{6c}: (0 0 1/2). The point group 
symmetry is \textit{-3m} (or $D_{3d}$). Thus, we also have
$\sigma_{xx}=\sigma_{yy}$.

\subsection{Calculation of the isotropic electric quadrupole cross-section at the Co \textit{K}-edge}
The space group symmetry operations of the group
$R\overline{3}m$ that leave the Co site 
(0,0,1/2) invariant are
$(x,y,z)$, $(-y,x-y,z)$, $(-x+y,-x,z)$, $(y,x,-z)$,
$(x-y,-y,-z)$, $(-x,-x+y,-z)$, $(-x,-y,-z)$, $(y,-x+y,-z)$,
$(x-y,x,-z)$, $(-y,-x,z)$, $(-x+y,y,z)$ and $(x,x-y,z)$.

The general angular dependence of electric quadrupole
transitions in terms of a second-rank spherical tensor
$\sigma^Q(2,m)$ and a fourth-rank spherical tensor
$\sigma^Q(4,m)$
was given in ref.~\onlinecite{ChB90}.
The trigonal symmetry of the Co site implies that the only
non-zero component of $\sigma^Q(2,m)$ is $\sigma^Q(2,0)$,
which is real.
To determine the non-zero components of
$\sigma^Q(4,m)$
for the Co site, we apply the formula derived in
ref.~\onlinecite{BJBA}:
\begin{eqnarray*}
\sigma^Q(4,m) &=& 
\frac{1}{12} \sum_{R m'} T^4_{m'} D^4_{m'm}(R),
\end{eqnarray*}
where $R$ runs over the rotation part of the 12 operations 
listed above and where 
$T^4_{m'}$ is an Hermitian fourth-rank tensor (without symmetry).
This gives us the only non-zero components of 
$\sigma^Q(4,m)$, which are 
$\sigma^Q(4,0)$ (a real number) and
$\sigma^Q(4,\pm 3 )$ (a purely imaginary number).
We define the real $\sigma^{Qi}(4,3)$ by
$\sigma^{Q}(4,\pm 3 )=i\sigma^{Qi}(4,3)$.

In a reference frame where the $x$ axis (the $z$-axis,
respectively) is along
the $\mathbf{a}$ vector (the $\mathbf{c}$ vector, respectively)
of the hexagonal lattice, this gives us
the angular dependence
\begin{eqnarray*}
\sigma^Q &=& 
\sigma^Q(0,0) +
\sqrt{\frac{5}{14}} (3 \sin^2\theta \sin^2\psi-1)
\sigma^Q(2,0) 
\\&+&
\sqrt{\frac{1}{14}} (35\sin^2\theta\cos^2\theta\cos^2\psi +
    5\sin^2\theta\sin^2\psi - 4)
\\&&
\times \sigma^Q(4,0) 
\\&&
-\sqrt{10}\sin\theta
\big[(3\cos^2\theta-1)\sin\psi\cos\psi\cos 3\phi
\\&&
+ \cos\theta(2\cos^2\theta\cos^2\psi-1)\sin 3\phi\big] \sigma^{Qi}(4,3).
\end{eqnarray*}
This expression is similar but not identical with the one
given in ref.~\onlinecite{ChB90} for the symmetry $D_{3d}$.
The difference comes from the fact that the relation between
the Cartesian axes and the local symmetry axes is different in 
the two cases.

The isotropic quadrupole absorption is $\sigma^Q(0,0)$.
In general, we need to calculate four different orientations
to deduce the isotropic quadrupole spectrum.
However, we could obtain it from only three directions:
$\hat\epsilon_1=(-1/\sqrt{6},-1/\sqrt{2},1/\sqrt{3}),
\mathbf{k}_1=(1/\sqrt{6},-1/\sqrt{2},-1/\sqrt{3})$,
$\hat\epsilon_2=(0,1,0), \mathbf{k}_2=(-1,0,0)$ and
$\hat\epsilon_3=(0,1/\sqrt{2},-1/\sqrt{2}),
\mathbf{k}_3=(0,1/\sqrt{2},1/\sqrt{2})$.
The corresponding absorption cross-sections are
\begin{eqnarray*}
\sigma^Q_1 &=& \sigma^Q(0,0) + \frac{\sqrt{14}}{9} \sigma^Q(4,0),\\
\sigma^Q_2 &=& \sigma^Q(0,0) + \frac{1}{\sqrt{14}} \sigma^Q(4,0)
   +2 \sqrt{\frac{5}{14}} \sigma^Q(2,0),\\
\sigma^Q_3 &=& \sigma^Q(0,0) + \frac{19}{4\sqrt{14}} \sigma^Q(4,0)
   - \sqrt{\frac{5}{14}} \sigma^Q(2,0).
\end{eqnarray*}
The isotropic quadrupole spectrum is
\begin{eqnarray*}
\sigma^Q(0,0) &=& \frac{27\sigma^Q_1-4\sigma^Q_2-8\sigma^Q_3}{15}.
\end{eqnarray*}

\end{document}